# Learning of viscosity functions in rarefied gas flows with physics-informed neural networks

Jean-Michel Tucny[a,b*], Mihir Durve[a], Andrea Montessori[b], Sauro Succi[a,c]

[a]Center for Life Nano- & Neuro-Science, Fondazione Istituto Italiano di Technology (IIT), viale Regina Elena 295, 00161 Rome, Italy

[b]Dipartimento di Ingegneria Civile, Informatica e delle Tecnologie Aeronautiche, Università degli Studi Roma Tre, via Vito Volterra 62, Rome, 00146, Italy

[c]Department of Physics, Harvard University, 17 Oxford St. Cambridge, MA 02138, United States

*Corresponding author e-mail: jean-michel.tucny@polymtl.ca

**Abstract.** The prediction non-equilibrium transport phenomena in disordered media is a difficult problem for conventional numerical methods. An example of a challenging problem is the prediction of gas flow fields through porous media in the rarefied regime, where resolving the six-dimensional Boltzmann equation or its numerical approximations is computationally too demanding. Physics-informed neural networks (PINNs) have been recently proposed as an alternative to conventional numerical methods, but remain very close to the Boltzmann equation in terms of mathematical formulation. Furthermore, there has been no systematic study of neural network designs on the performance of PINNs. In this work, PINNs are employed to predict the velocity field of a rarefied gas flow in a slit at increasing Knudsen numbers according to a generalized Stokes phenomenological model using an effective viscosity function. We found that activation functions with limited smoothness result in orders of magnitude larger errors than infinitely differentiable functions and that the AdamW is by far the best optimizer for this inverse problem. The design was found to be robust from Knudsen numbers ranging from 0.1 to 10. Our findings stand as a first step towards the use of PINNs to investigate the dynamics of non-equilibrium flows in complex geometries.

1. Introduction

Recently, deep learning methods have been proposed to recover the solution of non-linear partial differential equations (PDEs) [1, 2]. Deep learning methods show great promise for solving inverse problems on general transport problems in disordered media, where boundary conditions (BCs) or transport properties are unknown but constitutive relationships and experimental or numerical data about phenomena in the domain is available. Rather than attempting to discretize PDEs directly to solve systems of discretized equations, physics-informed neural networks (PINNs) use automatic differentiation to define all derivatives of physical variables on a meshless grid [3]. The PDE itself is used in the PINN as a residual to a loss function, which is an objective to be minimized. Additional information on constitutive equations or on the BCs of the problem can be added seamlessly, which requires much less testing and programming than with conventional CFD codes. When BCs can be written explicitly, mathematical problems are usually called forward problems, while PINNs learning relying mainly on data in the domain interior are usually called inverse problems.

A notoriously difficult physical problem to solve with conventional CFD methods is the study of porous media with gas flows in the rarefied regime, where the mean free path of gas molecules $\lambda$ is comparable to the characteristic length $L_C$ of the flow. In such rarefied gas flows, there is no clear separation of scales between the gradient of physical variables and momentum transfer at the molecular level. The Boltzmann equation (BE), describing the propagation and collision of population from a statistical mechanics point of view, is well known to be valid for all Knudsen numbers ($Kn = \lambda/L_C$) [4, 5]. Mesoscopic discretizations of the BE such as high-order Lattice-Boltzmann methods (LBM) [6] [7-13], discrete unified gas-kinetic schemes (DUGKS) [14-16] and discrete velocity methods (DVM) [17] have been shown to recover the Boltzmann equation for sufficiently large velocity sets. However, as the BE is six-dimensional, this process leads quickly to prohibitive computational and memory requirements. Furthermore, it is generally difficult to know *a priori* the required size of the velocity set as the Knudsen number increases. On one hand, microscopic representations such as the direct simulation Monte-Carlo (DSMC) method with its Lagrangian sampling of populations, is well suited for rarefied flows, but requires huge number of samples in the continuum limit to recover the gas flow in the continuum limit ($Kn \rightarrow 0$) [18, 19]. On the other hand, Eulerian macroscopic methods such as the Chapman-Enskog and Grad's moment methods study departure from equilibriums in the BE and attempt to recover closed forms for local macroscopic variables (density, velocity, temperature, etc.), and perform poorly as the Knudsen number increases [4, 20-23].

Inspired by such methods, PINNs have been used increasingly to solve rarefied gas flows. PINNs have been used to solve the Boltzmann-BGK equation in forward and inverse problems in the rarefied regime using a discrete-velocity method [24]. In the aforementioned paper, the researchers used a 3x3 velocity set for small Knudsen flows (D2Q9 lattice), while much larger velocity sets such as 28x28 velocity set is used for larger Knudsen flows both in forward and inverse problems. PINNs have also been developed using moment methods as a loss function in both forward and inverse problem formulations [25, 26]. However, the difficulty of knowing *a priori* the size of the velocity set required for the approximation of the BE remains in such formulations. PINNs have also been trained as inverse problem formulations on DSMC results to approximate the velocity field and their derivatives [27]. However, this approach might be prone to overfitting caused by a too large number of variables, especially with geometries of dimension larger than one.

An alternative to the use of BE approximations is a phenomenological description of rarefied gas flow phenomena with scaling laws for transport coefficients [28]. This approach shows great interest for finding solutions in industrial/engineering contexts, where we are not interested *per se* on the collisional and propagational behavior of populations, but rather on local macroscopic variables or integral variables such as flow rates, heat flows and work, etc. Two phenomena specific to rarefied gas flows which must be described as the Knudsen increases: 1) the apparent slip at the solid-gas boundary, and 2) departure from the linear shear-stress relationship. When $Kn > 0.001$, at the solid-gas boundary, gas molecules incoming from a region where the gas has a non-zero value creates an apparent slip flow. When $Kn > 0.1$, the Knudsen layer, a region of a thickness with $O \sim (\lambda)$, the Newtonian relationship between shear stress and shear strain is no longer valid as a result of insufficient gas-gas molecular collisions, takes a significant portion of the domain. Those physical phenomena are mathematically formulated in conventional CFD codes respectively as slip BCs on velocity and with an effective viscosity function in the Navier-Stokes solver. However, the coupling of those two phenomena may be difficult to obtain for disordered tridimensional geometries. To the authors' knowledge, PINNs were never used to recover the BCs and the effective viscosity for such an inverse problem. Moreover, the success of PINNs rely on the choice of activation functions, optimizers, neural structures, loss weights and other hyperparameters, for which research is very limited as stated in this recent review paper [29]. Such choices are generally written in research papers with little explanation and appear to be highly dependant both on the problem's physics and mathematical formulation. Studying PINN designs for this phenomenological approach is interesting for its applicability to other physical problems such as the Darcy flows [30], non-Newtonian flows for liquid polymers [31], and electrical conductivity in graphene [32].

In this paper, various implementations of a deep neural network informed by the generalized Stokes equation will be tested to recover the velocity field of a rarefied gas flow. The remainder of the article is divided as follows. In Section 2, the mathematical formulation of the generalized Stokes equation and data generation for the inverse problem will be presented. Various designs for the structure of the PINNs will be presented and tested respectively in Section 3 and 4. A robust, efficient and accurate PINN will be tested on different Knudsen numbers in Section 5. Concluding remarks will be related in Section 6.

## 2. The generalized Stokes model

In this section, we present the generalized Stokes model that will be leveraged inside the PINN to recover the velocity field of a rarefied gas flow. In Subsection 2a., the relationship between shear stress and shear strain, which breaks down in the Knudsen layer of a rarefied gas flow, will be defined. In subsection 2b., a velocity field through a slit benchmark will be derived to train the PINN later on.

### a. Definition of the effective viscosity

In this work, we consider a steady-state and incompressible gas flow moved by a uniform force field. In addition, we assume the flow to be laminar: i.e., the diffusive character of the flow is dominant. The Cauchy momentum equation can thus be written as follows [33]:

$$0 = -\frac{1}{\rho} \nabla \cdot \boldsymbol{\sigma} + \boldsymbol{f} \tag{1}$$

where $\rho$ is the density, $\sigma$ is the viscous stress tensor and $\mathbf{f}$ is a volumic force field (assumed in this work to be uniform). According to the kinetic theory of gases, the viscosity $\mu$ of a gas unbounded by solid walls can be derived from the thermodynamical properties of the gas as follows [4]:

$$\mu = \frac{\lambda P}{\sqrt{\frac{\pi \mathcal{R} T}{2}}} \tag{2}$$

where $P$ is the gas pressure, $T$ is the temperature of the gas and $\mathcal{R}$ is the specific perfect gas constant. In the Knudsen layer however, a region of a thickness of the order of the mean free path, insufficient collisions happen between gas molecules, which reduce the transfer of momentum. Therefore, viscosity, as the proportionality constant between the viscous stress tensor and the shear rate, can no longer be considered uniform. In the generalized Stokes equation, we assume the viscous stress tensor can be modeled with a space-dependant effective viscosity scalar field $\mu_e(\mathbf{x}) < \mu$ as follows [34]:

$$\boldsymbol{\sigma} = -\mu_e(\mathbf{x})\nabla \mathbf{v} \tag{3}$$

where $\mathbf{v}$ is the velocity field. In the general case, the effective viscosity is unknown a priori and should depend on the mean free path and the solid-fluid boundary. It is convenient to define the viscosity function $\Psi(\mathbf{x})$ as the ratio between the effective viscosity as follows:

$$\Psi(\mathbf{x}) = \frac{\mu_e(\mathbf{x})}{\mu} \tag{4}$$

For a rarefied gas flow in the laminar regime with negligible heat transfer, the following property is verified by construction: $0 < \Psi(\mathbf{x}) < 1$.

b. **Current benchmark: rarefied gas flow through a slit**

While the aforementioned formulation is a quantitatively accurate phenomenological description for planar flows, deriving and computing the viscosity function *a priori* is difficult and with questionable validity depending on the geometry, especially if the gas domain is non-convex [35, 36]. However, finding the solution to the Boltzmann equation is a problem of its own, and numerical methodologies require extensive verification and validation work. In this paper, we limit ourselves to the benchmark of a one-dimensional rarefied gas flow through a slit of size $D$, as shown in Fig. 1.

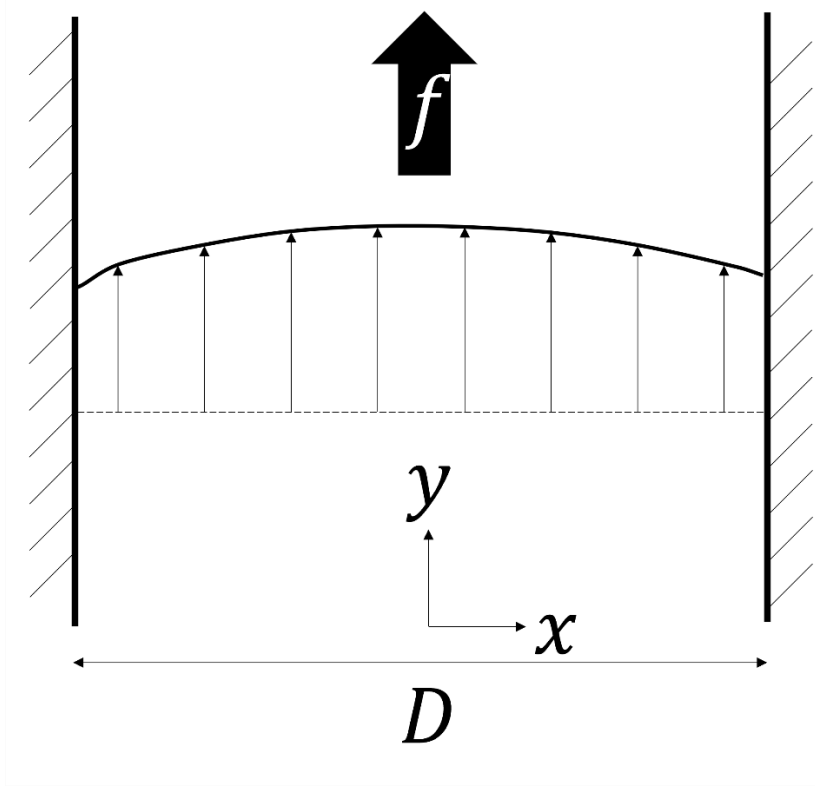

Fig. 1 – Schematic representation of a rarefied gas flow in a slit of size $D$ driven by a volumic force $f$.

It is proposed to generate the data with the phenomenological model proposed by Guo et al., where the unidimensional flow field generated through this model had been compared with DSMC numerical solutions [37] linearized Boltzmann equation solutions [38] and validated with experimental values. The viscosity function is defined as [34]:

$$\Psi(x) = \frac{1}{2}\left(F\left(\frac{\frac{D}{2}-x}{\lambda}\right) + F\left(\frac{\frac{D}{2}+x}{\lambda}\right)\right) \tag{5}$$

where $F(\beta)$ is defined as:

$$F(\beta) = 1 + (\beta - 1)\exp(-\beta) - \beta^2 E_i(\beta) \tag{6}$$

and where the exponential integral function $E_i(\beta)$ is defined as:

$$E_i(\beta) = \int_1^\infty t^{-1} \exp(-\beta t)\, dt \tag{7}$$

The following slip boundary condition from Guo is imposed [39]:

$$v_w = A_1 \lambda \Psi(x^*) \frac{\partial v}{\partial n} - A_2 \lambda^2 \Psi(x^*) \left(\frac{\partial}{\partial n}\left(\Psi(x^*) \frac{\partial v}{\partial n}\right)\right) \tag{8}$$

where $x^*$ is the location of a BC and is either equal to $-\frac{D}{2}$ or $\frac{D}{2}$ for the slit shown in Fig. 1, $n$ is the vector normal to the solid surface, $A_1 = 0.8183$ and $A_2 = 0.6531$ are phenomenological constants derived that can be derived from theoretical models or experimental values. Similar phenomenological models were validated for one-dimensional rarefied gas flows in the following articles [39-43].

The solution to Eq. (1) using the constitutive relationships in Eqs. (3), definitions of the viscosity function in Eqs. (4) and (5) and the BC (8) yields the following closed-form solution for the velocity:

$$v = -\frac{f}{\mu} \int_{-D/2}^{D/2} \frac{x}{\Psi(x)} dx \qquad (9)$$

The velocity data was generated through a trapezoid integration method with 4096 points. While the effective viscosity field (Eq. (5)) is generated at each point for the purpose of generating the velocity data, the effective viscosity field will only be used later for verification of the accuracy of the PINN, and not to train the PINN.

## 3. Physics-informed neural network designs

While the rarefied gas flow through a slit may be described using the generalized Stokes model and a second-order velocity BC, such a model may become ineffective for more complicated geometries. This warrants the training of PINNs to recover the velocity field without; 1) knowledge of a relationship between the velocity and other moments at the gas-solid interface; 2) knowledge of the viscosity function throughout the domain. In this section, we expose PINN concepts applied to this inverse problem, as well as various choices of implementations that are generally not systematically tested for the prediction of physical phenomena.

The most classical PINN architecture used in the literature for the approximation of PDEs for physical phenomena is the fully-connected PINN, which is shown in Fig. 2. The fully-connected PINN is made of $h$ hidden layers of $n$ neurons, takes as an input the position in the gas. Each layer of the fully-connected PINN expresses its output $\mathbf{z}^{(k)}$ as follows:

$$\begin{aligned} \mathbf{z}^{(0)} &= \mathbf{x} \\ \mathbf{z}^{(k)} &= \sigma\big(\mathbf{W}^{(k)}\mathbf{z}^{(k-1)} + \mathbf{b}^{(k)}\big) \\ \mathbf{z}^{(h)} &= \sigma\big(\mathbf{W}^{(h)}\mathbf{z}^{(h-1)} + \mathbf{b}^{(h)}\big) \end{aligned} \qquad (10)$$

where $\sigma(t)$ is a non-linear activation function, and $\mathbf{W}^{(k)}$ and $\mathbf{b}^{(k)}$ are the weight matrix and the bias vector of the $k$th layer ($k = (1, \cdots, h-1)$), which are the trainable model parameters. For the fully-connected PINN, the last layer $\mathbf{z}^{(h)}$ recovers both the velocity and the viscosity function. An example of a partially-connected PINN is shown in Fig. 3, where two sub-neural networks are created to compute the velocity and the viscosity function. Table 1 gives a summary of popular activation functions and their properties.

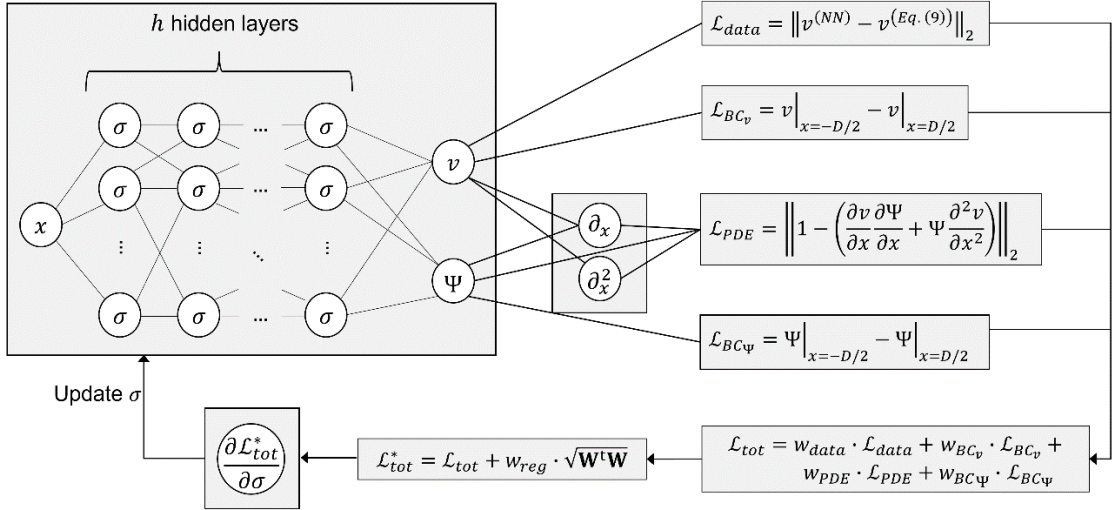

Fig. 2 – Depiction of a fully-connected PINN architecture to solve the extended Stokes inverse problem.

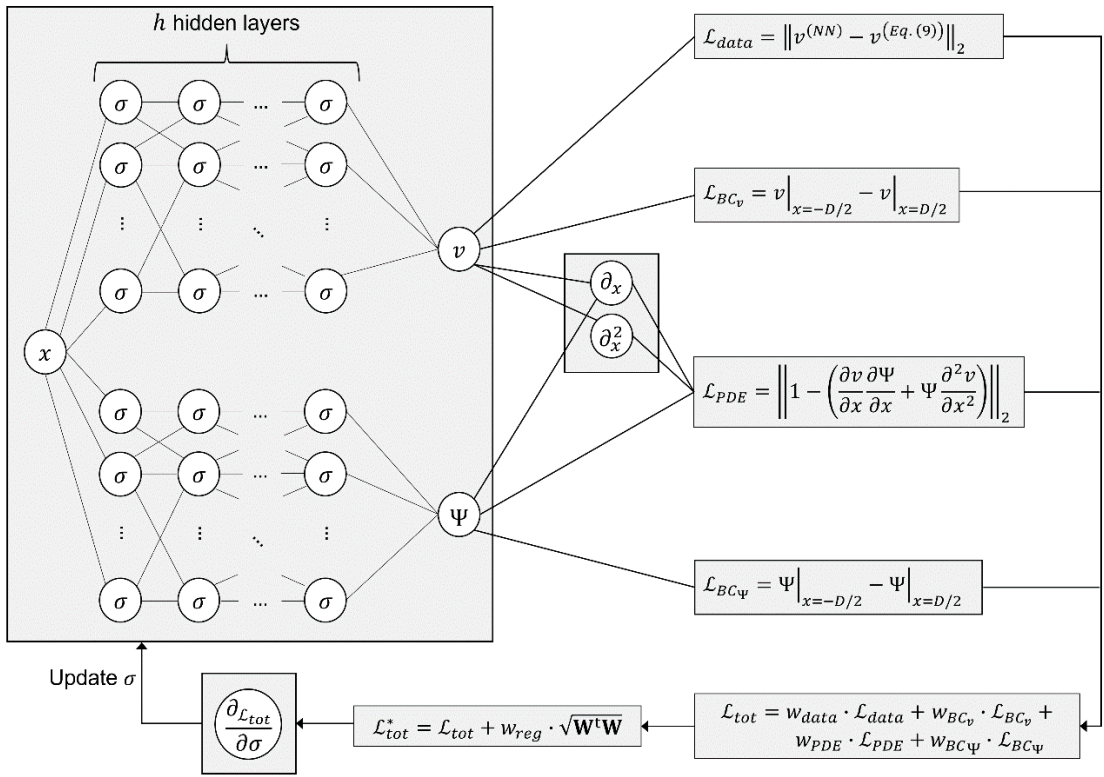

Fig. 3 – Depiction of a partially-connected PINN architecture to solve the extended Stokes inverse problem.

Table 1 – Description of activation function $\sigma$ and mathematical properties

| Name | $\sigma_t$ | Range | Smoothness |
|---|---|---|---|
| ELU [44] | $\begin{cases}(\exp(t)-1), & t \leq 0 \\ t, & t > 0\end{cases}$ | $(-1, \infty)$ | $C^1$ |
| ReLU [45] | $\max(0, t)$ | $(0, \infty)$ | $C^0$ |
| SeLU [46] | $\Omega\begin{cases}\alpha(\exp(t)-1), & t \leq 0 \\ t, & t > 0\end{cases}$ with parameters $\Omega \approx 1.0507$ and $\alpha \approx 1.6733$ | $(-\Omega\alpha, \infty)$ | $C^0$ |
| Sigmoid | $\dfrac{1}{1+\exp(-t)}$ | $(0,1)$ | $C^\infty$ |
| SiLU [47] | $\dfrac{t}{1+\exp(-t)}$ | $(-0.278, \infty)$ | $C^\infty$ |
| Tanh | $\tanh(t) = \dfrac{\exp(t) - \exp(-t)}{\exp(t) + \exp(-t)}$ | $(-1,1)$ | $C^\infty$ |

For both PINN architectures, it is aimed to minimize the total loss function $\mathcal{L}_{tot}$, which corresponds to a weighted sum of several loss terms. For this inverse problem, the distance $\mathcal{L}_{data}$ between the velocity predicted by the model and the velocity values produced with Eq. (9) is of primordial importance. The distinguishing factor between a statistical regression using deep learning and a PINN, however, is the definition of the PDE loss function $\mathcal{L}_{PDE}$, which is the residual corresponding to a normalized Eq. (1) and (3). The first-order and second-order partial derivatives related to velocities and viscosity in $\mathcal{L}_{PDE}$ are computed using automatic differentiation, which relies on accumulation of numerical values computed directly at code execution at machine precision [48]. All those loss functions are computed using the $L^2$ norm, which penalizes large local errors. In future configurations for rarefied gas flows, it might be possible to add a periodic boundary condition on all physical variables of interests, which gives additional constraints to the problem.

The weights and biases that will minimize the total loss function are computed using an optimizer. The basic principle works as follows: weights and biases are updated using automatic differentiation in successions of forward- and/or back-propagations through stochastic gradient descent methods. Famous examples include Root Mean Square Propagation (RMSProp [49]) and Adaptative Moment Estimation (Adam) [50]. Quasi-Newton methods such as the limited-memory Broyden-Fletcher-Goldfarb-Shanno (L-BFGS-B) algorithm [51] can also be used after an initial use of stochastic gradient descent. Derivatives of activation function at each node as a function of the PINN's weights and biases are again computed using automatic differentiation. The rate at which those weights and biases are modified by the optimizer is an hyperparameter called the learning rate.

To prevent overfitting of the neural network to the values and improve its robustness, $L^2$ regularization and weight decay contributions can also be supplemented to the PINN. $L^2$ regularization adds the $L^2$ norm of the weights to the total loss function:

$$\mathcal{L}_{tot}^* = \mathcal{L}_{tot} + w_{reg}\sqrt{\mathbf{W}^t\mathbf{W}} \qquad (11)$$

where $w_{reg}$ is the regularization hyperparameter. Alternatively, a weight decay term can be added to weights in the optimizer, the contribution of which is weighted with a weight decay hyperparameter $w_{wd}$. While the $L^2$ regularization and weight decay are similar in implementation for the RMSprop optimizer, researchers have suggested it is not the case for the Adam optimizer, and have proposed a new implementation called Adamw where the weight decay is properly set on the weights of the neurons [52].

As several hyperparameters are present in these PINNs, it will be easier to understand the meaning of the hyperparameters if all physical variables used to the train the PINN are normalized. In this study, the body force term $f$ and viscosity $\mu$ were set equal to one. This choice is valid because of the laminarity assumptions of our problem (or by inspecting Eq. (9)).

## 4. Test of PINN designs

In this section, the manifold of PINN designs for the approximation of the velocity and the viscosity field will be explored. As a full factorial design would be too fastidious to simulate and interpret, results are presented from the most critical to the least critical to the stability of the PINN solution, i.e.: 1) activation function and associated initializers, 2) optimizers and learning rates, 3) weights in the loss function, 4) PINN architectures, 5) size of the mini-batches. Preliminary trials have shown that the "base" PINN design shown in Table 2 is a good compromise of robustness and low computational requirements as a starting point for this parametric study, in the sense that varying one parameter from this base case allows it to converge and thus show more interesting results. However, this combination is not necessarily the one that gives the lowest prediction error. Note also that preliminary tests have not shown any significant difference between the Hammersley and other pseudorandom sequences.

Table 2 – "Base" PINN design for the parametric tests

| Activation function | Tanh |
|---|---|
| Optimizer | AdamW |
| Learning rate | 1E-5 |
| Weight decay | 1E-3 |
| Architecture | Fully-connected |
| Number of layers | 10 |
| Number of neurons per layer | 80 |
| Ratio of PDE to velocity residue weights | 1E-3 |
| Ratio of periodic BC to velocity residue weights | 1E-3 |
| Size of mini-batches | 32 points |
| Iterations per epoch | 100 |
| Choice of points in the mini-batch | Hammersley sequence |

Unless stated otherwise, the Knudsen number is always chosen to be $Kn = 1$, which is well into the transition regime and has interlapping Knudsen layers in the whole domain, but where gas-gas molecular collisions still give a significant gradient to the velocity and viscosity profile. All computations were performed on one CPU (Intel I9 64 GB Ram) and one GPU (Nvidia RTX3090, 24 GB of Ram, 10496 CUDA cores) using the DeepXDE [53] library with PyTorch v.1.13.1 [54] and Cuda v.11.7 [55].

### a. Choice of the activation function

As the activation function determines the relationship between the inputs and the outputs of each neuron, its properties are expected to be of prime importance in the convergence and accuracy of the neural network. The $L^2$ relative error of neural network predictions compared to the solution of Eq. (9) are presented in Table 3 for each activation function and its corresponding initialization as suggested by the literature, while velocity and viscosity functions are compared respectively with Eqs. (9) and (5) in Figs. 4-9.

It can be readily seen that activation functions with limited smoothness and non-compact ranges such as ELU, ReLU and SELU create orders of magnitude larger errors than the others on velocity, while being the main data furnished to the network. It can be noted that most of the spurious behavior of the velocity field in Figs. 4-6 a) appears in the middle of the slit, where the gradient of both the velocity and the viscosity functions are equal to zero. As derivatives of weights and biases are computed backwards in the automatic differentiation procedure, the learning process uses the inverse of the derivative of the loss function that is ill-conditioned when an activation function with limited smoothness is used. The neural network is unable to incorporate even a minute fraction of the physics-informed part in the loss function, especially considering the low value of the weight on the PDE residue compared to the weight on the data residue used in those tests ($\frac{w_{PDE}}{w_{data}} = 0.001$). In particular, according to the generalized Stokes model of a rarefied gas flow, the viscosity function should never be negative. Figs. 7-9, where results for infinitely differentiable functions are shown, show much better behavior.

In Fig. 7, it appears that after 2000 epochs, PINNs using the Sigmoid function is finally able to converge to Eqs. (9) and (5): however, convergence is rather slow as evidenced by its behavior at 200 epochs. While PINN outputs for the infinitely smooth SiLU function (shown in Fig. 8) behave better for the velocity profile, the PINN struggles to learn the viscosity function even after 2000 epochs. The Tanh activation function achieves the lowest relative errors and exhibit better convergence behavior, as further evidenced in Fig. 10. For this reason, the Tanh activation function will be retained for the remainder of the article. One can notice that the PINN easily learns the viscosity function in the center of the profile, with the largest error being committed at the edges of the domain. The magnitude of the error still remains lower than 2% despite: 1) the viscosity only appearing in the PDE term and 2) the lack of constraints on the viscosity function.

Table 3 – Performance of activation functions

| Activation function | Initialization | $L^2$ relative error on $v$ after 200 epochs (-) | $L^2$ relative error on $\Psi$ after 200 epochs (-) | $L^2$ relative error on $v$ after 2000 epochs (-) | $L^2$ relative error on $\Psi$ after 2000 epochs (-) | Runtime (s) |
|---|---|---|---|---|---|---|
| ELU | He normal | 2.32E-0 | 1.06E-0 | 2.33E-1 | 2.85E-0 | 1527 |
| ReLU | He normal | 2.72E-2 | 5.71E-1 | 1.82E-2 | 7.32E-1 | 1125 |
| SeLU | He normal | 2.86E-0 | 9.54E-1 | 1.36E-1 | 4.60E-0 | 1524 |
| Sigmoid | Xavier normal | 9.04E-2 | 2.82E-0 | 4.43E-4 | 1.92E-2 | 1386 |
| SiLU | He normal | 5.89E-4 | 1.24E-1 | 3.62E-4 | 3.17E-2 | 1634 |
| Tanh | Xavier normal | 6.20E-4 | 5.04E-2 | 5.03E-5 | 3.16E-3 | 1361 |

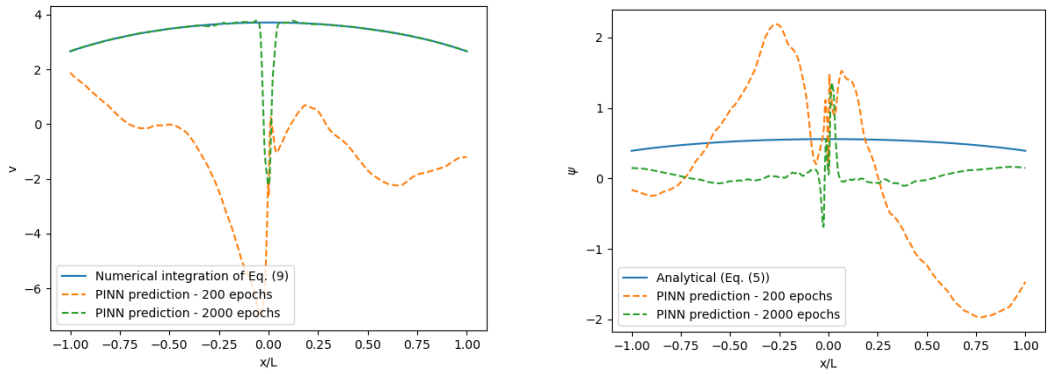

Fig. 4 – PINN results using the ELU activation function and the He normal initialization for a) the velocity $v$; b) the viscosity function $\Psi$

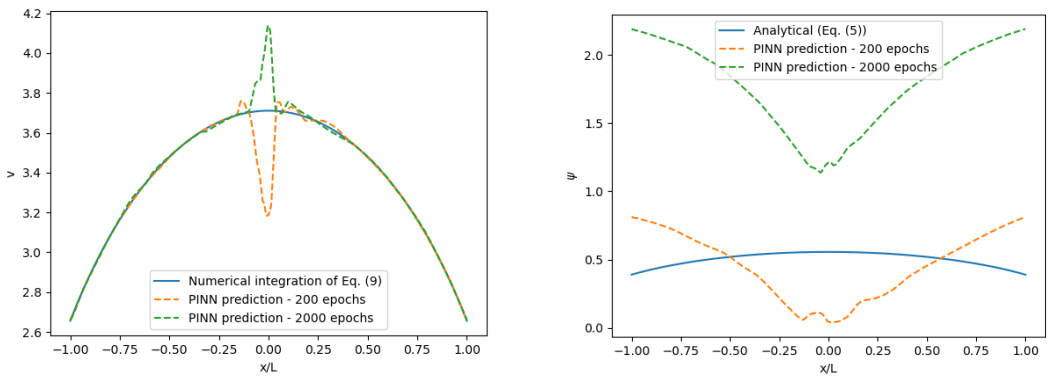

Fig. 5 – PINN results using the ReLU activation function and the He normal initialization for a) the velocity $v$; b) the viscosity function $\Psi$

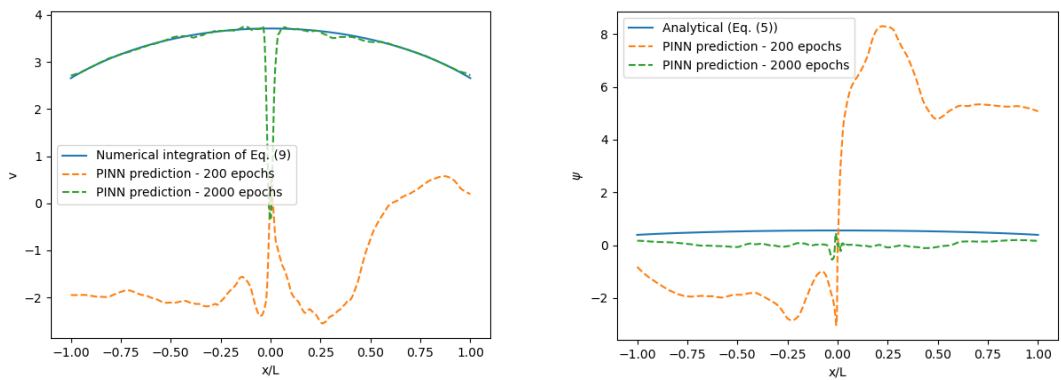

Fig. 6 – PINN results using the SeLU activation function and the He normal initialization for a) the velocity $v$; b) the viscosity function $\Psi$

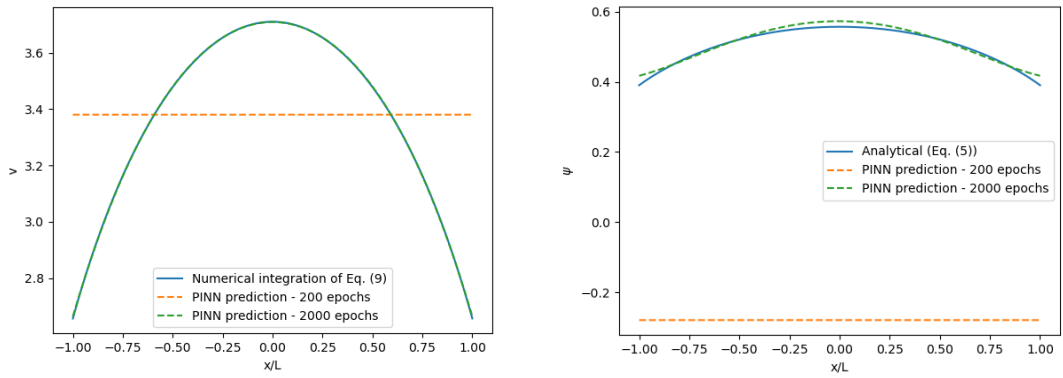

Fig. 7 – PINN results using the Sigmoid activation function and the Xavier normal initialization for a) the velocity $v$ and b) the viscosity function $\Psi$

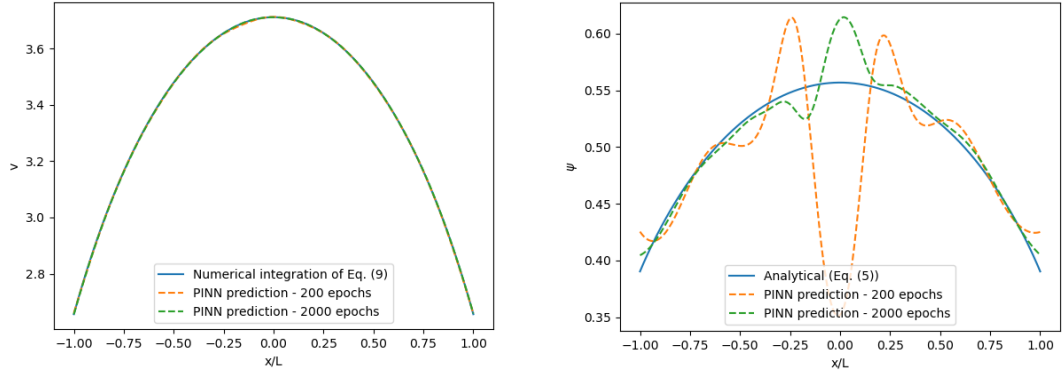

Fig. 8 – PINN results using the SiLU activation function and the He normal initialization for a) the velocity $v$ and b) the viscosity function $\Psi$

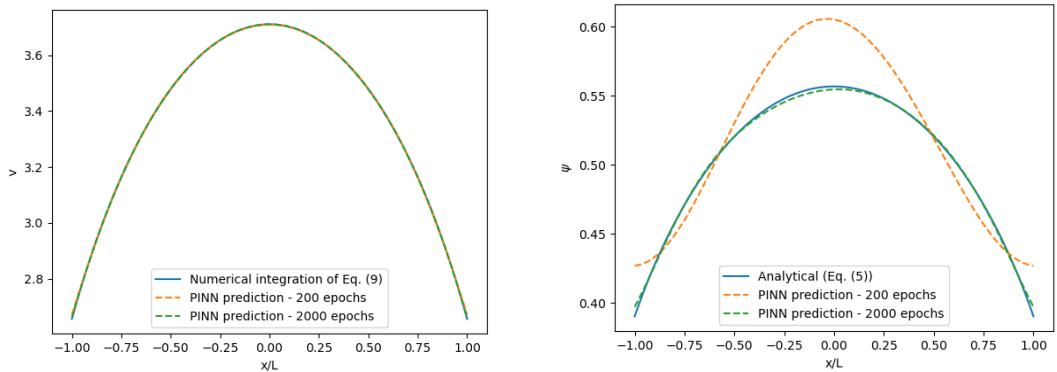

Fig. 9 – PINN results using the Tanh activation function and the Xavier normal initialization for a) the velocity $v$ and b) the viscosity function $\Psi$

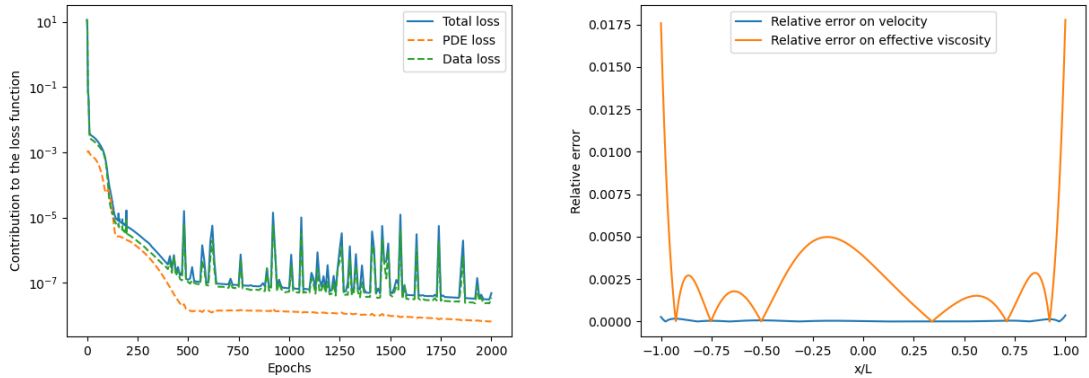

Fig. 10 – Evaluation of the a) Loss during training, and b) relative error after 2000 epochs for the PINN using the Tanh activation function and the Xavier normal initialization

### b. Choice of optimizers and learning rates

The performance of PINNs trained with common choices of optimizers (Adam+L2 regularization, AdamW and RMSProp are presented respectively in Tables 3-5. The use of the Adam optimizer + L2 regularization and the AdamW optimizer with a weight decay of zero is equivalent to the Adam optimizer: results for the Adam optimizers were thus added to Tables 3 and 4 for this reason for comparison purposes.

In Table 3, for the Adam optimizer with L2 regularization, the weight decay has a negative impact on the $L^2$ relative error both for the velocity field and the viscosity function. This could be expected, as the use of an additional constraint on the weight values decreases the relative impact of the velocity error on the loss function. However, in Tables 4 and 5, using a small weight decay of 1E-5 tends to slightly improve learning with the AdamW optimizer and RMSProp. Clearly, the AdamW with a weight decay of 1E-3 offers the best performance, with an order of magnitude smaller $L^2$ relative error with a slight 10% higher cost compared to the RMSProp optimizer.

Figs. 11-13 show PINN outputs of the Adam optimizer with L2 regularization, the AdamW and the RMSProp optimizers respectively for combinations of a 1E-5 weight decay and a 1E-6 learning rate. For neural networks converging with a low relative error in the end, this typical behavior is shown in Figs. 12-13. Although the velocity profile is qualitatively good after only 200 epochs, the viscosity function seems to have a curvature of opposite sign to the analytical solution. In Eqs. (5) and (9) one can remark both the effective viscosity and the velocity profiles are concave: i.e., the first derivatives are decreasing functions of $x$. However, further inspection of the PDE loss term of the PINN shows that first derivatives being increasing functions of $x$ if $\Psi$ increases or decreases accordingly, which corresponds to a saddle point of the loss function and is not easily avoided with Quasi-Newton optimization procedures such as the L-BFGS-B method. The use of the L-BFGS-B method has not led to significant improvement of any of the stochastic descent approaches (Adam+L2, AdamW and RMSProp).

Table 3 – Performance of the Adam optimizer + L2 regularization

| Weight decay | Learning rate | $L^2$ relative error on $v$ after 200 epochs (-) | $L^2$ relative error on $\Psi$ after 200 epochs (-) | $L^2$ relative error on $v$ after 2000 epochs (-) | $L^2$ relative error on $\Psi$ after 2000 epochs (-) | Runtime (s) |
|---|---|---|---|---|---|---|
| 1E-1 | 1E-3 | 1.01E-1 | 2.60E+4 | 1.01E-1 | 6.64E+3 | 1346 |
| 1E-1 | 1E-4 | 9.19E-2 | 2.05E+6 | 9.19E-2 | 1.56E+5 | 1348 |
| 1E-1 | 1E-5 | 9.01E-2 | 4.35E+0 | 9.00E-2 | 2.19E+6 | 1359 |
| 1E-1 | 1E-6 | 8.45E-2 | 2.48E+0 | 9.01E-2 | 6.42E+0 | 1358 |
| 1E-3 | 1E-3 | 6.65E-3 | 8.83E-2 | 6.19E-3 | 9.15E-2 | 1349 |
| 1E-3 | 1E-4 | 6.18E-3 | 8.99E-2 | 6.23E-3 | 1.00E-1 | 1356 |
| 1E-3 | 1E-5 | 4.96E-3 | 1.02E-1 | 4.41E-3 | 1.30E-1 | 1352 |
| 1E-3 | 1E-6 | 1.15E-2 | 3.50E-1 | 4.59E-3 | 5.78E-2 | 1352 |
| 1E-5 | 1E-3 | 2.35E-3 | 3.40E-2 | 7.05E-4 | 2.27E-2 | 1349 |
| 1E-5 | 1E-4 | 1.04E-3 | 3.27E-2 | 9.26E-4 | 2.70E-2 | 1361 |
| 1E-5 | 1E-5 | 8.37E-4 | 4.10E-2 | 5.49E-4 | 2.15E-2 | 1352 |
| 1E-5 | 1E-6 | 9.67E-3 | 2.28E-1 | 4.58E-4 | 2.15E-2 | 1358 |
| 0 | 1E-3 | 1.75E-3 | 6.12E-2 | 4.11E-4 | 1.91E-2 | 1336 |
| 0 | 1E-4 | 6.68E-4 | 2.07E-2 | 7.51E-4 | 2.43E-2 | 1342 |
| 0 | 1E-5 | 1.04E-3 | 4.87E-2 | 1.01E-4 | 3.68E-3 | 1345 |
| 0 | 1E-6 | 1.11E-2 | 2.26E-1 | 2.05E-4 | 7.36E-3 | 1348 |

Table 4 – Performance of the AdamW optimizer

| Weight decay | Learning rate | $L^2$ relative error on $v$ after 200 epochs (-) | $L^2$ relative error on $\Psi$ after 200 epochs (-) | $L^2$ relative error on $v$ after 2000 epochs (-) | $L^2$ relative error on $\Psi$ after 2000 epochs (-) | Runtime (s) |
|---|---|---|---|---|---|---|
| 1E-1 | 1E-3 | 3.91E-3 | 6.58E-2 | 7.23E-3 | 4.40E-1 | 1347 |
| 1E-1 | 1E-4 | 7.67E-4 | 1.97E-2 | 1.56E-4 | 1.08E-2 | 1343 |
| 1E-1 | 1E-5 | 5.60E-4 | 3.94E-2 | 9.98E-5 | 8.10E-3 | 1339 |
| 1E-1 | 1E-6 | 8.02E-3 | 2.38E-1 | 7.82E-5 | 4.90E-3 | 1356 |
| 1E-3 | 1E-3 | 4.66E-3 | 2.13E-2 | 1.59E-3 | 7.27E-2 | 1344 |
| 1E-3 | 1E-4 | 5.82E-4 | 2.15E-2 | 5.34E-4 | 2.07E-2 | 1341 |
| 1E-3 | 1E-5 | 5.91E-4 | 4.50E-2 | 5.80E-5 | 4.36E-3 | 1340 |
| 1E-3 | 1E-6 | 9.36E-3 | 2.80E-1 | 7.80E-5 | 4.48E-3 | 1345 |
| 1E-5 | 1E-3 | 1.04E-3 | 2.89E-2 | 1.47E-4 | 1.65E-2 | 1344 |
| 1E-5 | 1E-4 | 2.46E-4 | 1.04E-2 | 4.95E-4 | 8.49E-3 | 1341 |
| 1E-5 | 1E-5 | 5.84E-4 | 4.26E-2 | 4.12E-5 | 2.44E-3 | 1350 |
| 1E-5 | 1E-6 | 6.79E-3 | 2.24E-1 | 5.12E-5 | 3.32E-3 | 1350 |
| 0 | 1E-3 | 1.75E-3 | 6.12E-2 | 4.11E-4 | 1.91E-2 | 1336 |
| 0 | 1E-4 | 6.68E-4 | 2.07E-2 | 7.51E-4 | 2.43E-2 | 1342 |
| 0 | 1E-5 | 1.04E-3 | 4.87E-2 | 1.01E-4 | 3.68E-3 | 1345 |
| 0 | 1E-6 | 1.11E-2 | 2.26E-1 | 2.05E-4 | 7.36E-3 | 1348 |

Table 5 – Performance of the RMSProp optimizer

| Weight decay | Learning rate | $L^2$ relative error on $v$ after 200 epochs (-) | $L^2$ relative error on $\Psi$ after 200 epochs (-) | $L^2$ relative error on $v$ after 2000 epochs (-) | $L^2$ relative error on $\Psi$ after 2000 epochs (-) | Runtime (s) |
|---|---|---|---|---|---|---|
| 1E-1 | 1E-3 | 9.34E-2 | 2.05E+2 | 9.34E-2 | 2.06E+2 | 1200 |
| 1E-1 | 1E-4 | 9.26E-2 | 1.98E+3 | 9.37E-2 | 1.99E+3 | 1197 |
| 1E-1 | 1E-5 | 9.00E-2 | 3.76E+1 | 9.00E-2 | 6.40E+3 | 1200 |
| 1E-1 | 1E-6 | 8.46E-2 | 2.98E+0 | 9.11E-2 | 4.09E+0 | 1204 |
| 1E-3 | 1E-3 | 1.38E-2 | 1.51E+0 | 3.53E-3 | 3.98E-1 | 1203 |
| 1E-3 | 1E-4 | 1.12E-2 | 1.57E-1 | 9.34E-3 | 1.09E-1 | 1210 |
| 1E-3 | 1E-5 | 7.90E-3 | 1.27E-1 | 6.21E-3 | 9.12E-2 | 1204 |
| 1E-3 | 1E-6 | 1.14E-2 | 3.01E-1 | 6.10E-3 | 7.03E-2 | 1204 |
| 1E-5 | 1E-3 | 9.15E-2 | 2.18E+0 | 3.44E-3 | 3.65E-2 | 1215 |
| 1E-5 | 1E-4 | 5.13E-3 | 9.37E-2 | 3.63E-3 | 4.38E-2 | 1215 |
| 1E-5 | 1E-5 | 2.08E-3 | 5.92E-2 | 1.15E-3 | 2.58E-2 | 1214 |
| 1E-5 | 1E-6 | 6.29E-3 | 2.38E-1 | 6.14E-4 | 3.28E-2 | 1209 |
| 0 | 1E-3 | 1.83E-2 | 3.79E+0 | 1.95E-2 | 5.90E+0 | 1193 |
| 0 | 1E-4 | 7.62E-3 | 1.04E-1 | 1.50E-3 | 4.99E-2 | 1200 |
| 0 | 1E-5 | 3.16E-3 | 1.44E-1 | 2.34E-3 | 1.62E-2 | 1194 |
| 0 | 1E-6 | 2.43E-3 | 3.34E-1 | 2.70E-4 | 1.33E-2 | 1192 |

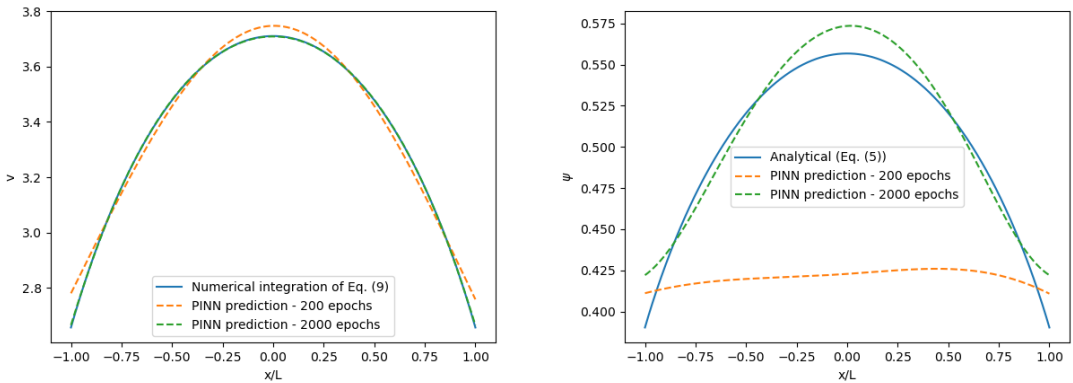

Fig. 11 - PINN results using the Adam optimizer + L2 regularization and a 1E-5 weight decay and a 1E-6 learning rate for a) the velocity $v$ and b) the viscosity function $\Psi$

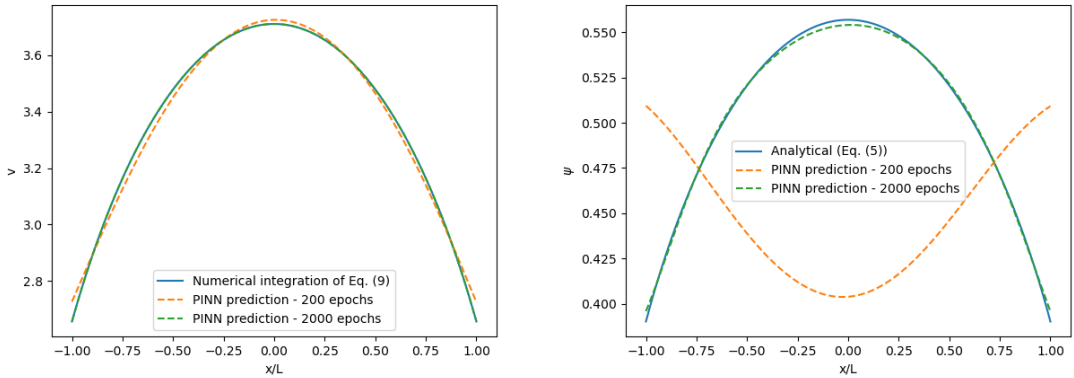

Fig. 12 – PINN results using the AdamW optimizer and a 1E-5 weight decay and a 1E-6 learning rate for a) the velocity $v$ and b) the viscosity function $\Psi$

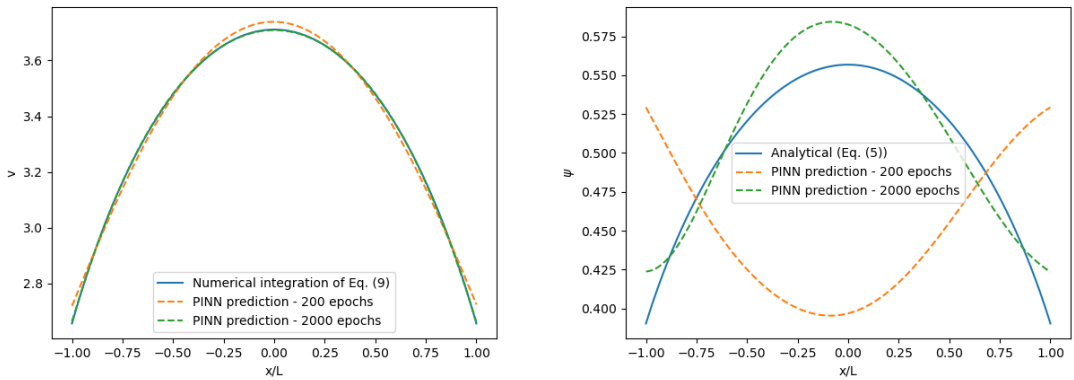

Fig. 13 – PINN results using the RMSProp optimizer and a 1E-5 weight decay and a 1E-6 learning rate for a) the velocity $v$ and b) the viscosity function $\Psi$

### c. Weights of the loss function

The ratio of the PDE loss weight to the data loss weight $\frac{w_{PDE}}{w_{data}}$ determines the importance of the "physics-informed" contribution compared to the data velocity contribution to the loss function. Table 6 shows the impact of $\frac{w_{PDE}}{w_{data}}$ on the PINN performance when no boundary conditions are imposed at the edges of the domain. The larger the contribution of the PDE loss weight, the lower the $L^2$ relative error. However, preliminary tests had shown that using a larger $\frac{w_{PDE}}{w_{data}}$ makes the PINN diverge more easily when using a larger learning rate or an alternative optimizer to AdamW.

Table 6 – Influence of the $\frac{w_{PDE}}{w_{data}}$ on the performance of a PINN without boundary condition

| Ratio of the PDE loss weight to the data loss weight $\frac{w_{PDE}}{w_{data}}$ | Learning rate | $L^2$ relative error on $v$ after 200 epochs (-) | $L^2$ relative error on $\Psi$ after 200 epochs (-) | $L^2$ relative error on $v$ after 2000 epochs (-) | $L^2$ relative error on $\Psi$ after 2000 epochs (-) | Runtime (s) |
|---|---|---|---|---|---|---|
| 1E-1 | 1E-3 | 5.42E-4 | 2.12E-2 | 3.50E-3 | 5.38E-2 | 1351 |
| 1E-1 | 1E-4 | 1.47E-3 | 3.37E-2 | 3.25E-3 | 1.01E-2 | 1346 |
| 1E-1 | 1E-5 | 1.05E-3 | 3.06E-2 | 5.16E-4 | 1.10E-2 | 1354 |
| 1E-1 | 1E-6 | 6.00E-3 | 1.66E-1 | 1.06E-4 | 4.91E-3 | 1354 |
| 1E-3 | 1E-3 | 1.36E-3 | 3.54E-2 | 7.14E-2 | 1.35E+0 | 1355 |
| 1E-3 | 1E-4 | 1.29E-3 | 3.77E-2 | 5.27E-4 | 1.45E-2 | 1348 |
| 1E-3 | 1E-5 | 2.40E-3 | 5.57E-2 | 1.28E-3 | 2.85E-2 | 1350 |
| 1E-3 | 1E-6 | 9.38E-3 | 2.18E-1 | 1.21E-4 | 4.87E-3 | 1357 |
| 1E-5 | 1E-3 | 2.64E-3 | 1.27E-1 | 9.02E-2 | 7.73E-1 | 1351 |
| 1E-5 | 1E-4 | 1.31E-3 | 4.37E-2 | 2.21E-3 | 7.23E-2 | 1357 |
| 1E-5 | 1E-5 | 8.24E-4 | 3.97E-2 | 3.65E-4 | 1.46E-2 | 1351 |
| 1E-5 | 1E-6 | 1.16E-2 | 8.51E-2 | 4.33E-4 | 1.21E-2 | 1353 |

It is difficult to have a general closed-form equation for a velocity boundary condition in rarefied gas flows. However, in the future, some structures might have a periodical geometry and this knowledge can be incorporate in the training of a PINN. As the solution in Eq. (9) is periodical in nature, the behavior of the PINN with a periodic BC can thus be tested. Fixing $w_{BC_\Psi} = w_{BC_v} = w_{PDE}$, PINNs were trained using this periodical boundary condition, for which performance results are shown in Table 7. For the slit benchmark, the imposition of periodic BCs does not appear to have much influence on $L^2$ relative error. However, this conclusion may have to be revisited for higher-dimensional problems in the future.

Table 7 - Influence of the $\frac{w_{PDE}}{w_{data}}$ on the performance of a PINN with $w_{BC_\Psi} = w_{BC_v} = w_{PDE}$

| Ratio of the PDE loss weight to the data loss weight $\frac{w_{PDE}}{w_{data}}$ | Learning rate | $L^2$ relative error on $v$ after 200 epochs (-) | $L^2$ relative error on $\Psi$ after 200 epochs (-) | $L^2$ relative error on $v$ after 2000 epochs (-) | $L^2$ relative error on $\Psi$ after 2000 epochs (-) | Runtime (s) |
|---|---|---|---|---|---|---|
| 1E-1 | 1E-3 | 1.55E-3 | 4.74E-2 | 6.01E-3 | 2.47E-2 | 1355 |
| 1E-1 | 1E-4 | 6.74E-4 | 2.95E-2 | 1.92E-4 | 8.03E-3 | 1350 |
| 1E-1 | 1E-5 | 9.34E-4 | 3.30E-2 | 3.09E-4 | 9.17E-3 | 1353 |
| 1E-1 | 1E-6 | 7.05E-3 | 2.07E-1 | 1.20E-4 | 6.11E-3 | 1354 |
| 1E-3 | 1E-3 | 3.48E-3 | 4.06E-2 | 3.16E-3 | 4.42E-2 | 1347 |
| 1E-3 | 1E-4 | 8.62E-4 | 4.47E-2 | 2.22E-4 | 9.72E-3 | 1355 |
| 1E-3 | 1E-5 | 9.62E-4 | 5.27E-2 | 1.78E-4 | 8.00E-3 | 1345 |
| 1E-3 | 1E-6 | 8.80E-3 | 1.98E-1 | 2.68E-4 | 8.78E-3 | 1350 |
| 1E-5 | 1E-3 | 5.78E-3 | 4.33E-1 | 9.01E-2 | 2.12E+1 | 1347 |
| 1E-5 | 1E-4 | 5.34E-4 | 9.13E-2 | 2.63E-4 | 2.77E-2 | 1347 |
| 1E-5 | 1E-5 | 1.01E-3 | 8.29E-2 | 1.86E-4 | 2.06E-2 | 1353 |
| 1E-5 | 1E-6 | 6.85E-3 | 2.07E+0 | 2.05E-4 | 2.67E-2 | 1345 |

### d. PINN architectures

In this subsection, the impact of the PINN architecture (a fully connected vs. a partially connected neural network as shown respectively in Figs. 2-3) is evaluated for various total numbers of neurons and numbers of neurons per layer. Tables 8 and 9 show respectively the performance of a fully connected PINN and of a partially connected PINN. For the fully connected PINN, the total number of neurons does not seem to have a substantial impact on the $L^2$ relative error and only contributes to larger runtimes. However, for a given total number of neurons, increasing the number of layers from 2 to 10 (i.e. making the neural network "deeper") substantially lowers the relative error found after 200 epochs both on the velocity and the viscosity function. However, the improvement of the PINNs with the number of layers is less systematic when the PINNs are left to converge for 2000 epochs. Similar conclusions are found for the partially connected PINN. Comparing the two architectures, it seems the partially connected PINN gives slightly smaller $L^2$ relative errors compared to the fully connected PINN. This is expected, noting however that preliminary tests have shown that partially connected PINNs diverge more easily if using a larger learning rate or an alternative optimizer to AdamW. It also seems the runtime is about 50% longer on the partially connected PINN than on the fully connected PINN. This result is somewhat surprising considering the lower number of gradients to be evaluated in the partially connected PINN. One explanation could be that designs with relatively low number of neurons do not fully take advantage of the acceleration potential of the GPU.

Table 8 – Performance of the fully-connected PINN

| Total number of neurons | Number of layers ($h$) | $L^2$ relative error on $v$ after 200 epochs (-) | $L^2$ relative error on $\Psi$ after 200 epochs (-) | $L^2$ relative error on $v$ after 2000 epochs (-) | $L^2$ relative error on $\Psi$ after 2000 epochs (-) | Runtime (s) |
|---|---|---|---|---|---|---|
| 200 | 2 | 9.90E-3 | 2.70E-1 | 2.08E-4 | 1.11E-2 | 376 |
| 200 | 4 | 7.32E-3 | 2.29E-1 | 3.65E-4 | 1.06E-2 | 569 |
| 200 | 5 | 2.22E-3 | 1.39E-1 | 2.67E-4 | 9.43E-3 | 635 |
| 200 | 10 | 1.53E-3 | 1.45E-1 | 3.43E-4 | 1.07E-2 | 1035 |
| 400 | 2 | 1.03E-2 | 2.85E-1 | 1.61E-4 | 9.73E-3 | 413 |
| 400 | 4 | 1.16E-3 | 3.83E-2 | 1.65E-4 | 8.58E-3 | 648 |
| 400 | 5 | 1.20E-3 | 6.22E-2 | 1.41E-4 | 7.43E-3 | 718 |
| 400 | 10 | 6.53E-4 | 3.51E-2 | 7.48E-5 | 4.21E-3 | 1114 |
| 800 | 2 | 8.96E-3 | 2.72E-1 | 2.26E-4 | 8.68E-3 | 604 |
| 800 | 4 | 1.06E-3 | 3.84E-2 | 9.73E-5 | 6.20E-3 | 811 |
| 800 | 5 | 2.89E-3 | 7.20E-2 | 5.89E-4 | 1.59E-2 | 888 |
| 800 | 10 | 7.05E-4 | 3.33E-2 | 1.60E-4 | 7.79E-3 | 1358 |
| 1600 | 2 | 9.32E-3 | 2.63E-1 | 7.82E-4 | 2.58E-2 | 1531 |
| 1600 | 4 | 7.60E-3 | 2.41E-1 | 1.39E-4 | 8.25E-3 | 1410 |
| 1600 | 5 | 4.98E-3 | 1.66E-1 | 7.61E-4 | 2.03E-2 | 1441 |
| 1600 | 10 | 1.80E-3 | 4.82E-2 | 8.02E-4 | 2.19E-2 | 1758 |

Table 9 – Performance of the partially-connected PINN with two subnetworks for the velocity and the viscosity function

| Total number of neurons | Number of neurons per layer ($h$) | $L^2$ relative error on $v$ after 200 epochs (-) | $L^2$ relative error on $\Psi$ after 200 epochs (-) | $L^2$ relative error on $v$ after 2000 epochs (-) | $L^2$ relative error on $\Psi$ after 2000 epochs (-) | Runtime (s) |
|---|---|---|---|---|---|---|
| 200 | 2 | 1.01E-2 | 2.73E-1 | 9.96E-5 | 6.27E-3 | 588 |
| 200 | 4 | 7.90E-3 | 2.50E-1 | 7.97E-5 | 3.54E-3 | 890 |
| 200 | 5 | 3.45E-3 | 2.33E-1 | 1.07E-4 | 7.14E-3 | 1076 |
| 200 | 10 | 2.54E-3 | 1.19E-1 | 2.08E-4 | 8.18E-3 | 1960 |
| 400 | 2 | 9.24E-3 | 2.78E-1 | 1.33E-4 | 8.36E-3 | 593 |
| 400 | 4 | 1.44E-3 | 6.37E-2 | 8.88E-5 | 5.54E-3 | 1008 |
| 400 | 5 | 1.32E-3 | 6.19E-2 | 1.52E-4 | 7.26E-3 | 1135 |
| 400 | 10 | 1.79E-3 | 7.40E-2 | 4.18E-4 | 1.31E-2 | 2067 |
| 800 | 2 | 9.60E-3 | 2.86E-1 | 1.87E-4 | 1.51E-2 | 660 |
| 800 | 4 | 1.05E-3 | 4.21E-2 | 7.97E-5 | 3.36E-3 | 1145 |
| 800 | 5 | 1.12E-3 | 4.61E-2 | 1.30E-4 | 9.06E-3 | 1375 |
| 800 | 10 | 2.17E-3 | 6.25E-2 | 5.53E-4 | 1.55E-2 | 2170 |
| 1600 | 2 | 8.85E-3 | 2.50E-1 | 2.88E-4 | 1.31E-2 | 1053 |
| 1600 | 4 | 7.12E-3 | 2.01E-1 | 1.52E-4 | 7.25E-3 | 1482 |
| 1600 | 5 | 2.00E-3 | 7.35E-2 | 6.85E-4 | 2.00E-2 | 1628 |
| 1600 | 10 | 8.21E-4 | 3.84E-2 | 1.98E-4 | 7.90E-3 | 2535 |

### e. Size of the mini-batch

The use of mini-batches for the PINN training allows the use of small subsets of the velocity data for the evaluation of the loss function. This feature would be especially useful for the training of PINNs on multidimensional flow problems. Table 10 shows the impact of the mini-batch size on PINN performance. As expected, the larger the mini-batch size, the larger the runtime and the smaller the $L^2$ relative error for all learning rates, especially for large learning rates. However, it seems that the $L^2$ relative errors tend to plateau for 64 data points or more.

Table 10 – Influence of the mini-batch size on PINN performance

| Number of data points in the mini-batch | Learning rate | $L^2$ relative error on $v$ after 200 epochs (-) | $L^2$ relative error on $\Psi$ after 200 epochs (-) | $L^2$ relative error on $v$ after 2000 epochs (-) | $L^2$ relative error on $\Psi$ after 2000 epochs (-) | Runtime (s) |
|---|---|---|---|---|---|---|
| 4 | 3 | 1.04E-2 | 2.18E-1 | 2.13E-2 | 6.08E+0 | 1087 |
| 4 | 4 | 3.75E-3 | 8.11E-2 | 6.04E-3 | 4.51E-1 | 1082 |
| 4 | 5 | 4.57E-3 | 7.05E-2 | 3.72E-3 | 5.14E-2 | 1081 |
| 4 | 6 | 7.76E-3 | 5.55E-1 | 2.21E-3 | 5.20E-2 | 1079 |
| 8 | 3 | 2.22E-2 | 3.23E-1 | 3.14E-2 | 6.90E+0 | 1135 |
| 8 | 4 | 4.97E-3 | 8.86E-2 | 4.88E-3 | 8.74E-2 | 1140 |
| 8 | 5 | 2.32E-3 | 5.56E-2 | 2.09E-3 | 4.68E-2 | 1138 |
| 8 | 6 | 1.13E-2 | 3.90E-1 | 1.03E-3 | 2.25E-2 | 1136 |
| 16 | 3 | 2.86E-3 | 4.57E-2 | 1.62E-3 | 4.18E-2 | 1198 |
| 16 | 4 | 9.27E-4 | 3.95E-2 | 3.37E-4 | 1.55E-2 | 1211 |
| 16 | 5 | 1.04E-3 | 4.54E-2 | 1.85E-4 | 1.10E-2 | 1206 |
| 16 | 6 | 1.09E-2 | 3.29E-1 | 1.08E-4 | 6.08E-3 | 1207 |
| 32 | 3 | 2.13E-3 | 5.17E-2 | 2.48E-3 | 4.93E-2 | 1348 |
| 32 | 4 | 9.50E-4 | 3.95E-2 | 6.10E-4 | 1.75E-2 | 1354 |
| 32 | 5 | 1.08E-3 | 3.88E-2 | 3.19E-4 | 1.14E-2 | 1345 |
| 32 | 6 | 1.08E-2 | 2.11E-1 | 8.80E-5 | 6.48E-3 | 1341 |
| 64 | 3 | 2.24E-3 | 7.21E-2 | 2.06E-3 | 4.89E-2 | 1548 |
| 64 | 4 | 4.12E-4 | 1.58E-2 | 2.89E-4 | 1.17E-2 | 1555 |
| 64 | 5 | 8.89E-4 | 4.57E-2 | 1.31E-4 | 7.37E-3 | 1551 |
| 64 | 6 | 8.56E-3 | 2.60E-1 | 5.44E-5 | 4.93E-3 | 1554 |
| 128 | 3 | 1.90E-3 | 5.43E-2 | 4.49E-3 | 6.38E-2 | 1898 |
| 128 | 4 | 5.81E-4 | 2.09E-2 | 3.07E-4 | 1.19E-2 | 1894 |
| 128 | 5 | 8.27E-4 | 4.55E-2 | 1.92E-4 | 9.42E-3 | 1904 |
| 128 | 6 | 8.21E-3 | 2.08E-1 | 6.83E-5 | 5.57E-3 | 1904 |
| 256 | 3 | 2.23E-3 | 5.16E-2 | 1.71E-3 | 7.85E-2 | 2703 |
| 256 | 4 | 6.68E-4 | 2.01E-2 | 1.96E-4 | 9.13E-3 | 2695 |
| 256 | 5 | 5.90E-4 | 3.65E-2 | 1.16E-4 | 5.99E-3 | 2710 |
| 256 | 6 | 6.97E-3 | 1.76E-1 | 6.71E-5 | 5.68E-3 | 2710 |
| 512 | 3 | 2.10E-3 | 5.20E-2 | 9.90E-4 | 3.44E-2 | 2700 |
| 512 | 4 | 1.21E-3 | 2.15E-2 | 2.25E-4 | 8.78E-3 | 2691 |
| 512 | 5 | 5.68E-4 | 4.33E-2 | 1.04E-4 | 7.05E-3 | 2705 |
| 512 | 6 | 7.58E-3 | 2.09E-1 | 5.89E-5 | 5.07E-3 | 2699 |

## 5. Influence of the Knudsen number on PINN performance

In this section, a PINN design based on the conclusions of Section 4 (summed up in Table 11) is tested for a gas flow through a slit on a wide range of Knudsen numbers. Figs. 14-17 show PINN results the velocity and viscosity function profiles while Table 12 shows the $L^2$ relative error for $Kn = 0.1, 0.5, 2$ and $10$ respectively. Despite widely different scales for the velocity and shapes for the effective viscosity profiles, the PINN was able to converge to results from the validated phenomenological model with a very low relative error after 2000 epochs. This shows the potential of PINNs to converge for rarefied gas flows using only the velocity field as training data and representation of the shear stress-shear strain relationship using an effective viscosity as a scalar field.

Table 11 – PINN design for the slit benchmark for the range $0.1 < Kn < 10$

| Activation function | Tanh |
|---|---|
| Optimizer | AdamW |
| Learning rate | 1E-5 |
| Weight decay | 1E-3 |
| Architecture | Partially connected |
| Number of layers | 4 |
| Number of neurons per layer | 100 |
| Total number of neurons | 800 |
| Ratio of PDE to velocity residue weights | 1E-3 |
| Ratio of periodic BC to velocity residue weights | 1E-1 |
| Size of mini-batches | 64 points |
| Iterations per epoch | 100 |
| Choice of points in the mini-batch | Hammersley sequence |

Table 12 – PINN performance for the slit benchmark for the range $0.1 < Kn < 10$

| Knudsen number | $L^2$ relative error on $v$ after 200 epochs (-) | $L^2$ relative error on $\Psi$ after 200 epochs (-) | $L^2$ relative error on $v$ after 2000 epochs (-) | $L^2$ relative error on $\Psi$ after 2000 epochs (-) | Runtime (s) |
|---|---|---|---|---|---|
| 0.1 | 5.26E-3 | 7.53E-2 | 5.63E-4 | 9.45E-3 | 1221 |
| 0.5 | 2.32E-3 | 4.87E-2 | 1.71E-4 | 6.18E-3 | 1220 |
| 2.0 | 9.04E-4 | 6.47E-2 | 2.56E-5 | 5.16E-3 | 1209 |
| 10.0 | 3.44E-1 | 9.93E-1 | 3.18E-5 | 6.43E-3 | 1237 |

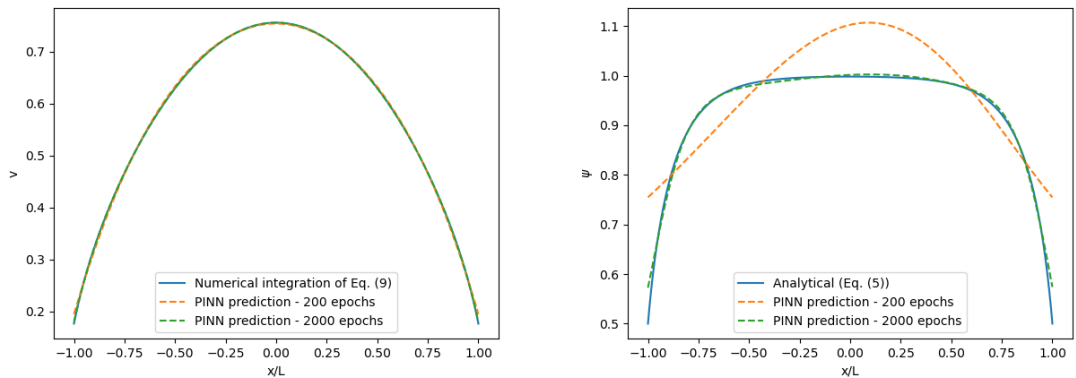

Fig. 14 – PINN results for the slit benchmark with a $Kn = 0.1$ with the profiles for: a) the velocity $v$ and b) the viscosity function $\Psi$

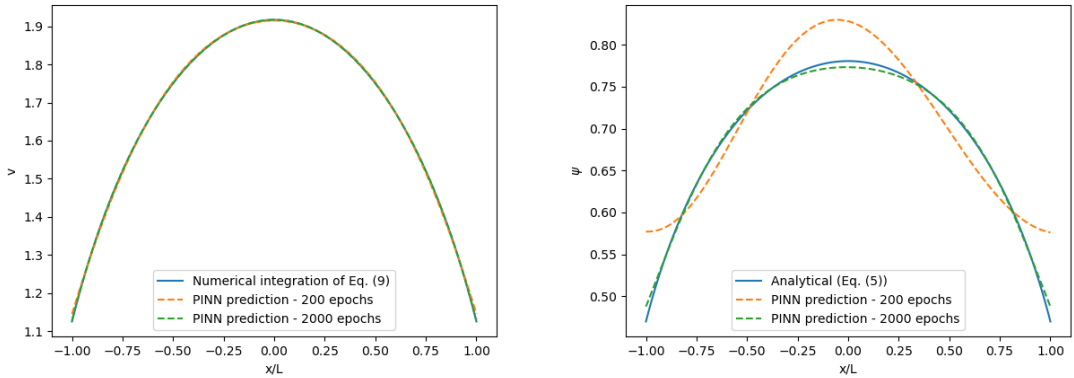

Fig. 15 – PINN results for the slit benchmark with a $Kn = 0.5$ with the profiles for: a) the velocity $v$ and b) the viscosity function $\Psi$

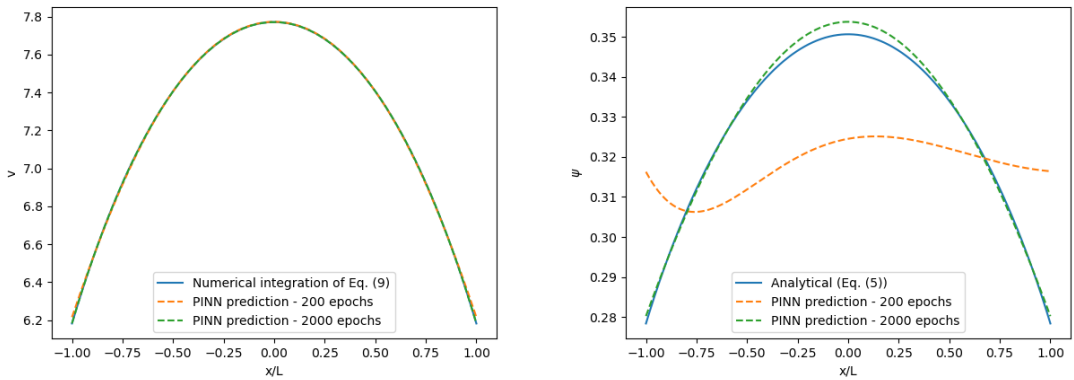

Fig. 16 - PINN results for the slit benchmark with a $Kn = 2$ with the profiles for: a) the velocity $v$ and b) the viscosity function $\Psi$

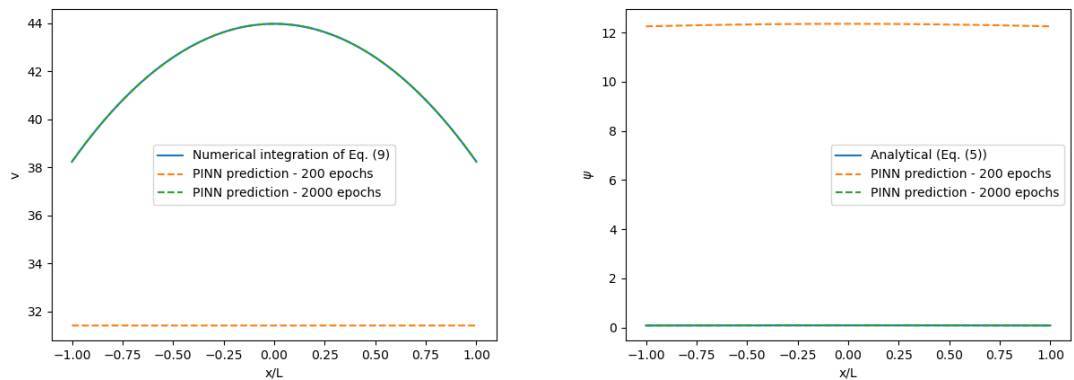

Fig. 17 - PINN results for the slit benchmark with a $Kn = 10$ with the profiles for: a) the velocity $v$ and b) the viscosity function $\Psi$

## 6. Conclusion

Characterization of non-equilibrium physical phenomena using a constitutive relationship is a well-known approach to model complex systems. In this paper, PINNs have been proposed to learn the effective viscosity of a rarefied gas flow through a slit. The only information that was given to the PINN in the loss function was the velocity data obtained from a validated phenomenological model and the local linearity of the effective viscosity function. More importantly, no explicit information was given on the velocity of the boundary condition or on the shape of the effective viscosity function, which makes it a challenging inverse problem for a PINN to be trained on.

As the manifold of PINN designs was not well documented in the literature, a large parametric study was made to find a PINN design that is both robust and accurate. It was found that a combination of the Tanh activation function and the AdamW optimizer is key to the convergence of the PINN for this problem. A relatively low 1E-3 ratio between the PDE loss weight and the velocity data loss weight should be used for robustness and accuracy. Periodic boundary conditions have not significantly improved the accuracy of the PINN. Counter-intuitively, a partially-connected PINN design with two sub-networks for the velocity and viscosity function slightly improved the result, while slightly increasing its runtime on the tested benchmark. The use of a mini-batch containing as low as 32 data points does not significantly impact the accuracy of the PINN while significantly improving the runtime. The accuracy of the PINNs was demonstrated on a wide range of Knudsen numbers $0.1 < Kn < 10$ exhibiting quite different shapes for the velocity and the effective viscosity functions.

The findings of this paper have potential to be applicable to any characterization of physical phenomena through a PINN using a constitutive relationship. However, all the PINN training was limited on a relatively simple one-dimensional slit geometry. Future work will include the test of our PINN on physical phenomena with higher-dimensional geometries.

**Acknowledgements**

J.-M. T. thanks the FRQNT "Fonds de recherche du Québec – Nature et technologies (FRQNT)'' for financial support (Research Scholarship No. 314328). The authors acknowledge funding from the European Research Council Grant Agreement No. 739964 (COPMAT) and ERC-PoC2 grant No. 101081171 (DropTrack).